\documentclass[useAMS,usenatbib,fleqn]{mn2e}
\usepackage{amsmath}
\usepackage{multirow}
\usepackage{graphicx}
\usepackage{epsfig}
\usepackage{amssymb,amsmath}
\usepackage{aas_macros}
\usepackage{fixltx2e} 
\DeclareMathOperator\erf{erf}

\title[Simulated EHT Experiments]{
The Event Horizon Telescope:  exploring strong gravity and accretion physics
}

\author[Ricarte \& Dexter]
 {Angelo Ricarte, $^1$ Jason Dexter $^2$ \\
 $^1$ Yale University, Department of Astronomy, J.W. Gibbs Laboratory, 260 Whitney Avenue, New Haven, CT, 06511, USA \\
 $^2$ Departments of Physics and Astronomy, University of California, Berkeley, CA 94720-3411, USA}

\date{\today}

\begin{document}
\pagerange{\pageref{firstpage}--\pageref{lastpage}} \pubyear{2014}
\maketitle

\label{firstpage}

\begin{abstract}

The Event Horizon Telescope (EHT), a global sub-millimeter wavelength very long baseline interferometry array, is now resolving the innermost regions around the supermassive black holes Sgr A* and M87. Using black hole images from both simple geometric models and relativistic magnetohydrodynamical accretion flow simulations, we perform a variety of experiments to assess the promise of the EHT for studying strong gravity and accretion physics during the stages of its development. We find that (1) the addition of the LMT and ALMA along with upgraded instrumentation in the ``Complete'' stage of the EHT allow detection of the photon ring, a signature of Kerr strong gravity, for predicted values of its total flux; (2) the inclusion of coherently averaged closure phases in our analysis dramatically improves the precision of even the current array, allowing (3) significantly tighter constraints on plausible accretion models and (4) detections of structural variability at the levels predicted by the models. While observations at 345 GHz circumvent problems due to interstellar electron scattering in line-of-sight to the galactic center, short baselines provided by CARMA and/or the LMT could be required in order to constrain the overall shape of the accretion flow. Given the systematic uncertainties in the underlying models, using the full complement of two observing frequencies (230 and 345 GHz) and sources (Sgr A* and M87) may be critical for achieving transformative science with the EHT experiment.

\end{abstract}

\section{Introduction}
The Event Horizon Telescope (EHT) is a very long baseline interferometry (VLBI) array consisting of radio telescopes around the world.\footnote{http://www.eventhorizontelescope.org}  Once complete, it will be capable of achieving a resolution of up to 23 microarcseconds at 230 GHz and 15 microarcseconds at 345 GHz.  This resolution is required, as its name suggests, to resolve black hole structure on event horizon scales.  Both the supermassive black hole at the center of the Milky Way, Sagittarius A*, and that of the elliptical galaxy M87 are massive enough and close enough (subtending tens of microarcseconds) for the EHT to resolve.  Sgr A* is predicted to subtend a larger angle on the sky, and with a declination of $\approx -29^{\circ}$, it remains permanently visible in the Southern Hemisphere.  However, radio images of Sgr A* are blurred due to interstellar scattering by turbulent electrons \citep{bac78,bow06}, which is consistent with a thin scattering screen located near the Scutum spiral arm \citep{bow14}.  Supergiant elliptical galaxy M87, known for its striking jet in the optical, provides an alternate source--it has a black hole nearly as large in angular size, and without any indication of strong interstellar scattering at 7mm \citep[e.g.,][]{had11}. Initial EHT observations have made the first detections of event horizon scale structure using both sources \citep{doe08,fis11,doe12}. The array capabilities are rapidly improving both in sensitivity and coverage due to increased bandwidth and the inclusion of additional radio telescopes \citep[e.g., the Atacama Large Millimeter/submillimeter Array (ALMA),][]{fis13}. 

Here we assess the potential for future EHT observations at various milestones in its development, taking into account the locations and sensitivities of the EHT's constituent telescopes.  We attempt to answer a variety of observational questions. Can the EHT detect signatures of strong field general relativity? How well can we constrain models for black hole images?  Should we preferentially observe Sgr A* or M87?  What are the tradeoffs between observing at 345 GHz versus 230 GHz? We perform a set of experiments using geometric and theoretical models for black hole images (\S\ref{sec:modeling}) and our simulated arrays (\S\ref{sec:arrays}). We statistically quantify the performance of each simulated observing run (\S\ref{sec:experiments}), and discuss the implications and limitations of this study (\S\ref{sec:discussion}) before summarizing our results (\S\ref{sec:conclusion}).

\section{Modeling}
\label{sec:modeling}

The sparse uv-coverage of the current EHT array prevents imaging \citep[but see][for an example of image formation with the future EHT array]{lu14}. Maximizing the science return on the data or simulating future observations requires fitting models in the Fourier domain. Both geometric shapes and physical emission models have been used to interpret the observations. The mm emission from Sgr A* and M87 is thought to be synchrotron radiation \citep[e.g.,][]{rey80,mel94,fal98}, either from a radiatively inefficient accretion flow \citep{nar95,yua03,rey96} or a jet \citep{fal00a,bro09}. Recently, there has been considerable interest in making predictions for the observed event horizon scale structure, both using semi-analytic \citep[e.g.,]{bro09a,yua09} and MHD \citep[e.g.,][]{mos09,dex09,shc12} accretion flow and jet models. Thus far both geometric and physical models can produce satisfactory fits \citep[e.g.,][]{bro09,dex10,fis11}, but physically-motivated ones are statistically favored \citep{bro11b,kam13}. In this work, we use both physical models from numerical simulations of the innermost regions of accretion flows and jets and a parameterized geometric description of their resulting ``crescent'' images.

\begin{figure*}
\begin{center}
\includegraphics[scale=0.8]{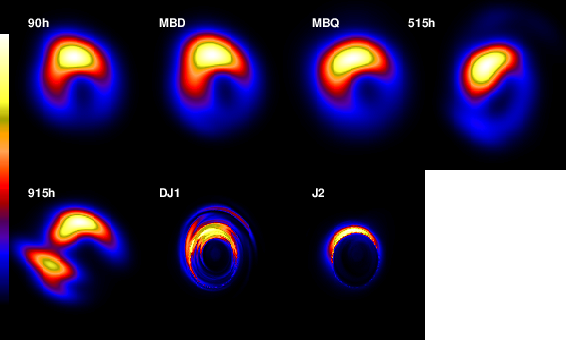}
\caption{False-color black hole images from numerical models of Sgr A* (top and 915h) and M87 (DJ1 and J2) used in our simulated EHT observations, shown on a linear scale. The properties of the images are described in Table \ref{table:sim_images}. The Sgr A* images are best fit models to existing EHT data, and have been scatter-broadened to mimic the blurring effects of interstellar scattering. Although calculated from a range of theoretical models, in all cases the images are roughly crescent-shaped due to the combined relativistic effects of Doppler beaming and gravitational light bending. Each image has been normalized to its maximum flux, and scaled via the color bar shown on the left. \label{fig:sim_images}}
\end{center}
\end{figure*}

\begin{table*} 
\begin{center}
\begin{small}
\begin{tabular}{lcccccc}
\hline
Name  & Source  &     a  &  i ($^\circ$) & Type &  Field  &  References (simulation, image)\\
\hline
90h   &     Sgr A*  & 0.9   & 60    &   Disk                 &               Poloidal     &     \citet{fra07,dex09,dex10}\\
MBD   &    Sgr A* &  0.92  &60   &    Disk                  &              Dipolar       &     \citet{mck09,dex10}\\
MBQ   &    Sgr A* &  0.94 & 50   &    Disk                    &            Quadrupolar  &    \citet{mck09,dex10}\\
515h   &   Sgr A* &  0.5  &  40  & Tilted Disk              &             Poloidal       &    \citet{fra09,dex13}\\
DJ1     &    M87   &   0.92 & 25   &    Disk/jet              &              Dipolar      &     \citet{mck09,dex12a}\\
J2       &    M87   &   0.92 &  25   &    Jet                     &             Dipolar        &    \citet{mck09,dex12a}\\
\hline
\end{tabular}
\end{small}
\end{center}
\caption{Properties of the black hole images from simulations\label{table:sim_images}}
\end{table*}

\subsection{Simulated Black Hole Images}

The theoretical black hole images used here are the result of relativistic radiative transfer calculations from MHD simulations of accretion flows \citep[e.g.,][]{dex12}. We use images corresponding to the global best fit to sub-mm Sgr A* data (spectral and EHT) both from aligned simulations \citep{fra07,mck09} and from those where the angular momentum of the accreting material was misaligned from the black hole spin axis by $15^\circ$ \citep{fra07,fra09}. More details about the calculations and fitting procedure can be found in \citet{dex10,dex12,dex13}. Some basic properties of the models we use are given in Table \ref{table:sim_images}, including the black hole spin $a$, inclination $i$, the initial magnetic field configuration, and the portion of the gas distribution which dominates the observed sub-mm emission. The images are shown in Figure \ref{fig:sim_images}. These simulated images are necessarily a small subset of those in the literature. However, as discussed below the predicted images are strongly shaped by relativistic effects, and so we believe the results of simulated EHT observations using these models will be similar to those using images calculated by other groups \citep[e.g.,][]{bro01,nob07,mos09,bro11b,shc12}. These are taken as a range of plausible models for Sgr A* and M87, but are not intended to span the true parameter space of possibilities for the outcome of EHT observations.

\subsection{Geometric Parameterization}

The theoretical models described thus far to calculate event horizon scale black hole images either from semi-analytic models or MHD simulations suffer from systematic uncertainties due to incomplete physics and uncertain initial conditions. For this reason, we also employ a geometric crescent (+ ring) model for the image of an accretion flow. The crescent is the difference between two constant intensity disks \citep{kam13}.  This model aims to capture the dominant relativistic effects near black holes: Doppler beaming and gravitational light bending. The accretion flow moving toward our line of sight is ``beamed'' to a higher flux, leading to asymmetry in the image, while light from the back side of the accretion flow is deflected above and below the black hole to the observer. These combined effects lead to a crescent image morphology \citep[e.g.,][and see Figure \ref{fig:sim_images}]{bro01}.

A geometric crescent model is defined by five parameters \citep{kam13}:  $R_{out}$, the radius of the outer edge of the crescent; $R_{in}/R_{out}$, the fractional radius of the hole in the crescent; $\phi$, the position angle, defined as the angle between the u-axis and the line between the center of the crescent's hole and the center of the crescent; $F_C$, the total flux of the crescent; and $\xi$, the displacement of the crescent's hole from the center, parameterized as a fraction of $R_{out} - R_{in} \in [0,1)$. The asymmetry parameter $\xi$ can be thought of as a proxy for the strength of Doppler beaming, and in the context of standard accretion flow models as a proxy for the inclination of the accretion disk with respect to our line of sight.  A face-on disk should have symmetric flux ($\xi \approx 0$, at least when averaged over long timescales), while an edge-on disk can be highly asymmetric from Doppler beaming ($\xi \rightarrow 1$). 

In addition to the accretion flow, captured by the crescent, general relativity predicts that black holes should cast a ``shadow'' corresponding to the region where photon orbits reaching the observer are bound to the black hole \citep{bar73,fal00b}. Outside of this location, optically thin images show a bright ring corresponding to the unstable circular photon orbit, located at a physical radius of $3 G M / c^2$.  Detection of this photon ring is tantamount to the discovery of the shadow. In addition, the size of the ring allows a measurement of the mass and distance of the black hole \citep{joh11}, while its shape can in principle be used to test the ``no-hair theorem'' of general relativity \citep{joh10}. Simulated models \citep{dex09,dex10,dex13} predict that this ring could contribute $\simeq1-10\%$ of the total flux, estimated by extracting the fraction of the flux concentrated just outside the location of the predicted location of the lensed photon orbit.

The ring component, when added, has only one parameter:  $F_R$, its total flux.  It is parameterized as a delta function fixed at a radius of $\sqrt{27} G M / c^2 \simeq 27 (M/4\times10^6 M_\odot) (8 \rm kpc / D) \, \mu as$, the lensed orbit for the ``photon ring'' of a non-spinning black hole. Although the photon ring becomes highly asymmetric at high values of black hole spin \citep{bar73}, its size only varies by $\simeq 10\%$.  This component is predicted by general relativity, but its existence has not yet been confirmed via observation.  Given its physical extent, we determine the angular size of this ring from the mass and distance of each of our target objects, whose basic parameters are listed in table \ref{table:basic_parameters}.  Note that while recent gas-dynamical studies of M87 suggest a black hole mass of $3.5^{+0.9}_{-0.7} \times 10^9 \, M_\odot$ \citep{wal13}, we use the larger value determined by stellar-dynamical studies.

\begin{table} 
\begin{center}
\resizebox{0.48\textwidth}{!}{
\begin{tabular}{ |c|c|c| }
\hline
Parameter & Sgr A* & M87 \\
\hline
$M_{BH}$ ($M_{\sun}$) & $4.31 \times 10^6$ & $6.6 \times 10^9$ \\
$D$ (kpc) & $8.33$ & $1.79 \times 10^4$ \\
Diameter of Shadow ($\mu$as) & 53.0 & 37.8 \\
Right Ascension & 17h45m40.03599s &  12h30m49.42338s \\
Declination & -29$^{\circ}$00'28.1699'' & +12$^{\circ}$23'28.0439'' \\
\hline
\end{tabular}
}
\end{center}
\caption{Assumed values for Sgr A* and M87 parameters.  Although the black hole at the center of M87 is three orders of magnitude farther, it is also three orders of magnitude more massive.  Masses and distances of Sgr A* and M87 were provided by \citet{gil09} and \citet{geb11} respectively.  Coordinates for Sgr A* were provided by \citet{pet11}, while those for M87 are from \citet{lam09}. For comparison, the lensed angular diameters of the shadows ($2 \cdot \sqrt{27} G M /c^2$) of these black holes are calculated.  \label{table:basic_parameters}}
\end{table}

The complete Fourier Transform of the crescent model, $\mathfrak{F}(u,v)$, where $u$ and $v$ are in units of $\lambda$, can be expressed analytically as:

\begin{center}
\begin{equation}
\mathfrak{F}(u,v) = G(u,v) \left[F_{C} C(u,v) + F_{R} R(u,v)\right], \label{eqn:total}
\end{equation}
\end{center}

\noindent where $C(u,v)$ and $R(u,v)$ are the Fourier transforms for a crescent and ring,

\begin{center}
\begin{align}
C(u,v) &= \frac{F_C}{\pi (R_{out}^2 - R_{in}^2)} \bigg[D(R_{out},\sqrt{u^2 + v^2})  \label{eqn:crescent}\\
&- D(R_{in},\sqrt{u^2+v^2})  \nonumber\\ 
&\cdot \exp(2 \pi i \xi (R_{out}-R_{in})(\cos(\phi)u+\sin(\phi)v))\bigg] \nonumber\\
R(R,\rho) = &J_0(2 \pi R \rho),
\end{align}
\end{center}

\noindent $D(R,\rho)$ is the Fourier transform of a constant intensity disk of size R at radial spatial frequency $\rho$,

\begin{center}
\begin{equation}
D(R,\rho) = J_1(2 \pi R \rho) R/\rho \label{eqn:disk},
\end{equation}
\end{center}

\noindent and $J_n (x)$ are Bessel functions of the first kind. Finally, $G(u,v)$ is the Fourier transform of the Gaussian, which accounts for the blurring effect of strong interstellar scattering towards Sgr A*, using empirical values from \citet{bow06}. 

\subsection{Making Measurements}
\label{sec:measurements}

The $(u,v)$ coordinates sampled by a given EHT array depend on the location of the telescopes, the time of year, and the position of the source.  The calculations to determine appropriate coordinates are carried out using the MIT Array Performance Simulator\footnote{http://www.haystack.mit.edu/ast/arrays/maps/} (MAPS), a program designed to simulate radio interferometry observations.  Each observing run is a 24 hour time period set on the April 5th of 2009, with 10 minute scans on-source beginning every 20 minutes.  (The actual date is unimportant, since we simulate observations for the full 24 hour period and do not consider the effects of the Sun or other celestial objects in our analysis.)  We assume 10 second integrations, which are limited by the atmosphere coherence time.  Each telescope is assumed to have elevation limits of between 15 and 85 degrees, with the exception of PdBI, which can observe down to 10 degrees.  Gaussian noise is applied to the real and imaginary components of these measurements, characterized by a standard deviation of $\sqrt{SEFD_{1} SEFD_{2}/2 B \tau}$, where $SEFD_{1,2}$ is the System Equivalent Flux Density of each telescope in the baseline, $B$ is the bandwidth, and $\tau$ is the integration time \citep{tho01}. The observables we consider, visibility amplitudes and closure phases, are insensitive to atmospheric phase delays which we ignore.

\subsubsection{Real, Imaginary, and Amplitudes}
EHT data thus far have almost exclusively consisted of visibility amplitudes. However, most of the information in an image is encoded in its phase.  One peculiarity of the absence of phase information is a $180^\circ$ degeneracy due to the fact that $|\mathcal{F}(u,v)| = |\mathcal{F}(-u,-v)|$. Unfortunately, atmospheric phase delays make it challenging to record Fourier phases using VLBI. Hence, we often simulate measurements comprised only of visibility amplitudes by taking the magnitude of simulated real and imaginary visibilities.  Such a measurement is assigned a Gaussian error bar equal to the \emph{same} error quoted above in the context of visibilities, which follows from linear error propagation theory.  Amplitude measurements have approximately Gaussian errors, so normal $\chi^2$ statistics can be applied to their measurements. The appropriate errors in closure phase measurements are described below.

\subsubsection{Closure Phases}
\label{sec:closure_phases}
While atmospheric phase delays make measurements of Fourier phases nearly impossible, {\it closure phases} of Sgr A*, the sum of phases between three different telescopes, have been successfully measured \citep{fis11} and will constitute a crucial EHT observable. Although closure phases do not include all of the full Fourier phase information, they are insensitive to all site-specific errors such as atmospheric and hardware delays \citep{rog76}.  Because of this, closure phases can be coherently averaged for much longer than the atmospheric coherence time.  

A straightforward linear Gaussian error propagation exercise would imply that the error on each phase should be $1/s$, where $s$ is the signal-to-noise ratio, and that these errors could be added in quadrature when measuring a closure phase.  Yet in the limit where $s$ is small, phase noise is non-Gaussian, and we notice discrepancies from  Gaussian statistics.  For simulated observations of amplitudes and closure phases, we calculate amplitudes and their errors as above, then assign an error on the phase measurement based on the probability distribution for phase noise \citep{rog84}.  

\begin{center}
\begin{align}
p(\theta) &= \frac{e^{-s^2/2}}{2 \pi} \bigg[ 1 + \bigg(\frac{\pi}{2} \bigg)^{1/2} s \cos{\theta} e^{s^2\cos^2(\theta)/2} \label{eqn:phase_error}\\
                &\cdot \bigg( 1 + \erf\bigg(\frac{s \cos{\theta}}{\sqrt{2}} \bigg) \bigg) \bigg] \nonumber\\
\end{align}
\end{center}

\noindent In the limit of large $s$, this converges to

\begin{center}
\begin{equation}
p(\theta) \approx \frac{s e^{-s^2\theta^2/2}}{(2 \pi)^{1/2}} \label{eqn:phase_error_limit},
\end{equation}
\end{center}

\noindent a normal distribution with a standard deviation from linear Gaussian error propagation formulas. Note that we have corrected errors found in both of these equations in our source, \citet{rog84}.

To minimize computational time, we calculate only a minimal set of the \emph{independent} closure phases, selecting those which can measured with the greatest precision.  Hence, for each array, we designate the telescope with the minimum SEFD as the fixed telescope with which all closure triangles are made.  Then, for each simultaneous measurement, phases are summed for each triangle containing the fixed telescope.  The effective signal-to-noise ratios of the resultant closure phases are calculated using equation \ref{eqn:closure_phase_error}.  This is equal to the {\it inverse} of the fractional closure phase uncertainty.   

\begin{center}
\begin{equation}
S_{cp}  = L^{-1}[L(s_1)L(s_2)L(s_3)] \label{eqn:closure_phase_error}
\end{equation}
\end{center}
where $L$ is given by
\begin{center}
\begin{equation}
L(s) = s\left(\frac{\pi}{8}\right)^{1/2}e^{-s^2/4}[I_0(s^2/4)+I_1(s^2/4)]
\end{equation}
\end{center}
and $I_0$ and $I_1$ are hyperbolic Bessel functions.

In the high signal-to-noise limit, equation \ref{eqn:closure_phase_error} asymptotes to 

\begin{center}
\begin{equation}
S_{cp} \approx (s_1^{-1/2} + s_2^{-1/2} + s_3^{-1/2})^{-1/2}  \label{eqn:closure_phase_error_linear},
\end{equation}
\end{center}

\noindent which would be the familiar result from linear Gaussian error propagation theory.

Because closure phases can be coherently averaged for much longer than the atmospheric coherence time ($\tau = 10$ s), we allow them to be averaged for the the full integration time, $T = 10$ minutes, as in \citet{bro11a}.  This has the effect of shrinking the error bars on closure phases by a factor of $\sqrt{T/\tau} = \sqrt{60}$ under our assumptions.  In general, $T$ is limited by the inherent variability of the source (unknown a priori).

\subsection{Fitting}
\label{sec:fitting}

Using this modeling procedure, we can generate best fit static ringed and ringless crescents to 230 GHz observations of Sgr A* \citep{doe08,fis11} and M87 \citep{doe12}. These fits provide us a template for the truth images used in several of our experiments. Best fit parameters and uncertainties are calculated using the MCMC fitting algorithm described in \citet{goo10}.  Our MCMC ensemble is comprised of 100 walkers and is run through 1000 trials ($10^5$ total samples). The initial configuration of the walkers is a Gaussian distribution in parameter space, instantiated with realistic values and widths corresponding roughly to the uncertainties in each parameter (as determined by previous runs).  We verify that the our results are insensitive to these initial conditions.  The tunable acceptance parameter, which \citet{goo10} denote as $a$, is set to 3 in order to force the acceptance ratio below 50\%.  

Convergence is determined by watching the means of the parameter distributions, as well as the average $\chi^2$ asymptote to a stable value.  While convergence is typically seen within the first $\sim10$\% of the MCMC, we conservatively discard the first 50\% to ensure that our samples are distributed according to the true posterior PDF.  We organize the remainder of the models into bins in each parameter, which span between $\pm 4$ standard deviations from the parameter's mean.  The number of bins is given by $8(N/24 \pi)^{1/3}$, where $N$ is the number of models included in the histogram.  This is the optimal number of bins spanning this width estimated from bias-variance theory, assuming an inherently normal distribution.\footnote{The Bias-Variance Tradeoff applied to histogram bins is given by $\Delta x = \frac{1}{n^{1/3}} \left(\frac{6}{\int(f'(x))^2 dx}\right)^{1/3}$, where $\Delta x$ is the bin width, $n$ is the number of data points, and $f'(x)$ is the first derivative of the PDF.  The formula we apply is obtained by assuming $f(x) = \frac{1}{\sqrt{2 \pi} \sigma}\exp{-\frac{(x-\mu)^2}{2 \sigma^2}}$, where $\mu$ is the peak of the PDF, and $\sigma$ is its standard deviation, estimated in the usual manner.}   The histogram created this way defines each parameter's marginalized posterior probability distribution.  We define the ``best fit'' parameters to be the modes of these distributions (i.e., the values with the highest probability) when binned as above.  In order to obtain uncertainties in these values, we determine the width in parameter space which encloses 68.3\% (1 $\sigma$) of the models on either side of the mode.  For parameters without strict boundary conditions, these distributions tend to be roughly Gaussian.

\begin{table*}
\begin{center}
\resizebox{\textwidth}{!}{
\begin{tabular}{ |c|c||cccccc| }
\hline
Object
   & Model
   & $R_{out}$ ($\mu$as)
   & $R_{in}/R_{out}$
   & $\phi$ ($^{\circ}$)
   & $F_C$ (Jy)
   & $\xi$
   & $F_R$ (Jy) \\
\hline
Sgr A*
   & Gaussian
   & $40^{+3}_{-3}$
   & $0.39^{+0.03}_{-0.02}$
   & $110.2^{+1.4}_{-1.3}$
   & $2.41^{+0.04}_{-0.03}$
   & $-$
   & $-$ \\
Sgr A*
   & Crescent
   & $26.2^{+3.7}_{-0.9}$
   & $0.95^{+0.03}_{-0.23}$
   & $69^{+4}_{-4}$
   & $2.229^{+0.024}_{-0.019}$
   & $0.93^{+0.05}_{-0.11}$
   & $-$ \\
Sgr A*
   & Crescent + Ring
   & $27.0^{+1.9}_{-1.2}$
   & $0.89^{+0.07}_{-0.10}$
   & $66^{+3}_{-5}$
   & $2.2^{+0.06}_{-0.21}$
   & $0.97^{+0.02}_{-0.09}$
   & $0.06^{+0.22}_{-0.04}$ \\
M87
   & Gaussian
   & $17.3^{+0.4}_{-0.3}$
   & $0.5^{+0.2}_{-0.3}$
   & $14^{+5}_{-7}$
   & $0.958^{+0.014}_{-0.009}$
   & $-$
   & $-$ \\
M87
   & Crescent
   & $63.3^{+0.9}_{-1.4}$
   & $0.78^{+0.03}_{-0.03}$
   & $68^{+5}_{-8}$
   & $1.301^{+0.018}_{-0.022}$
   & $0.992^{+0.005}_{-0.061}$
   & $-$ \\
\hline
\end{tabular}
}
\end{center}
\caption{Best fit parameters to 230 GHz Sgr A* and M87 data. $\xi$ is close to 1 in each of these fits, indicating that a significant offset of the hole from the center of the crescent is favored.  For Gaussian fits, $R_{out}$ corresponds to the major axis, $R_{in}$ corresponds to the minor axis, $\phi$ corresponds to the angle with which the major axis is tilted from the horizontal, and $F_C$ corresponds to the total flux.  Due to the nature of visibilities with poor (u,v) coverage, fitting different parameterizations can yield significantly different structures as seen in the M87 cases.  \label{table:best_fit_parameters}}
\end{table*}

Best fit parameters for the crescent and Gaussian models are listed in Table \ref{table:best_fit_parameters}, and figure \ref{fig:crescent_model} displays the images corresponding to these parameters after accounting for electron scattering.  Our ringless crescent and Gaussian model parameters are consistent with previous work \citep{bro11b,kam13}. For Sgr A*, the inclusion of the photon ring causes only a slight shift in the best fit crescent parameters. The inferred flux in the ring component is poorly constrained, but is marginally non-zero and consistent with the fraction found in theoretical models of Sgr A* (see below). The inclusion of the photon ring is not statistically supported by M87 data (see \ref{sec:photon_ring}). 

\begin{figure*}
\begin{center}$
\begin{array}{cc}
\includegraphics[width=0.5\textwidth]{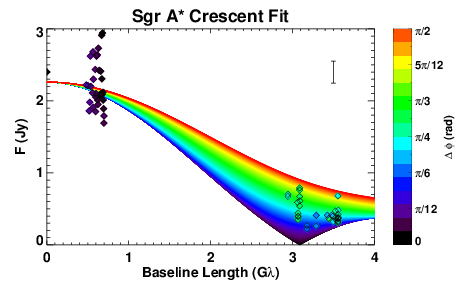} & 
\includegraphics[width=0.5\textwidth]{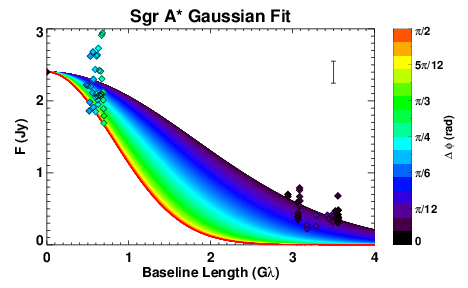} \\
\includegraphics[width=0.5\textwidth]{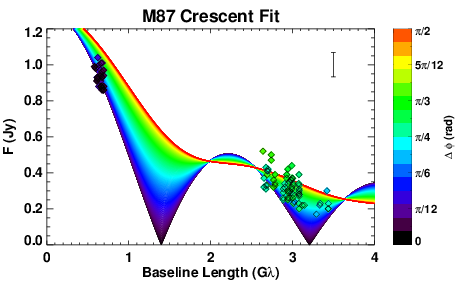} &
\includegraphics[width=0.5\textwidth]{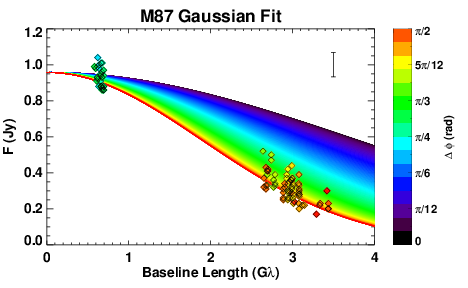} \\
\end{array}$
\caption{Visibility amplitudes measured by the EHT overlaid with our best-fit models.  Each color represents a slice through a different angle within the Fourier plane. Recorded data points are color-coded according to the angle in the Fourier plane to which they apply, and a typical error bar is shown in the top right.  Model degeneracies are apparent, which would be resolved with a better sampling of Fourier space.   
\label{fig:model_amplitudes}}
\end{center}
\end{figure*}

Current EHT data from both sources are insufficient to distinguish between these different models, and this is further illustrated in figure \ref{fig:model_amplitudes}, a visualization of our fits.  Here, EHT visibility amplitudes are overplotted on the visibility amplitudes of our best fit models given in \ref{table:best_fit_parameters}.  The x-axis represents radius in the Fourier plane, while each color from purple to red represents a slice at a different angle, sampled evenly in the interval $[\phi+\pi/2, \phi+\pi)$, where $\phi$ is the position angle of the model.  (Apart from the subtle asymmetry of scattering at these values, these models are symmetric about the line $\theta = \phi+\pi$, in addition to the usual rotational symmetry in the (u,v) plane.)  Data points are color coded according to the section of the surface they are meant to fit. 

These fits reveal the dangers of overfitting data poorly-sampled in Fourier space, as qualitatively different models can produce similar visibility amplitudes at the uv locations measured by current stations.  The total flux of M87 can vary noticeably between models, since our M87 data lack an independent flux constraint.  Note further that the M87 best-fit crescent has a much larger size than its best-fit Gaussian, made possible because the long baselines in the crescent fit measure flux from a higher-order peak in the power spectrum.  In general, such subtleties make it challenging to find global best-fit parameters, as many completely different models can produce similar fits. These degeneracies will be broken with the inclusion of phase information, which discriminate between adjacent peaks of the power spectrum, and additional telescopes, which improve (u,v) coverage and more precisely measure its shape.

\section{Arrays}
\label{sec:arrays}

In the context of VLBI, an ``array'' corresponds to an instantiation of the the entire EHT, while a ``telescope'' corresponds to a single element of the EHT, which is sometimes an interferometer that has been phased together to function as a single dish. To represent the EHT at different stages of its development, we devise two sets of increasingly complete arrays.  One of these sets operates at 230 GHz, for which there is existing data and a greater number of telescopes.  The other set operates at 345 GHz, a frequency which is less affected by the astrophysical noise noted in section \ref{sec:modeling}.  The telescopes and their System Equivalent Flux Densities (SEFD) for each of these arrays \citep{doe09} are displayed in Table \ref{table:arrays}.  The current 230 GHz array uses a bandwidth of 1024 MHz, while all other arrays use a bandwidth of 4096 MHz, a bandwidth which is soon to be achieved. 

\begin{table*} 
\begin{center}
\begin{tabular}{ |c|c|c|c|c|c|c|c| }
\cline{2-8}
\multicolumn{1}{r}{}  & \multicolumn{4}{ |c| }{230 GHz} & \multicolumn{3}{ |c| }{345 GHz} \\ \hline
Telescopes & Current (0) & Precision (1) & Coverage (2) & Complete (3) &  Basic (0) & Coverage (1) & Future (2) \\ \hline
JCMT & 9400 & & & & & & \\ 
SMTO & 11900 & 11900 & 11900 & 11900 & 23100 & 23100 & 23100 \\ 
CARMA & 17700 & 3500 & 3500 & 3500 & & & 4900 \\ 
SMA & & 4900 & 4900 & 4900 & 8100 & 8100 & 8100 \\ 
LMT & & & 10000 & 560 & & & \\ 
APEX & & & 6500 & 6500 & 12200 & 12200 & 12200 \\ 
ASTE & & & & &14000 & 14000 & 14000 \\ 
ALMA & & & & 110 & & 1000 & 1000 \\ 
PV & & & & 2900 & & & \\ 
PdBI & & & & 1600 & & 3400 & 3400 \\ 
SPT & & & & 7300 & & & \\ \hline 
Bandwidth (MHz) & 1024 & 4096 & 4096 & 4096 & 4096 & 4096 & 4096 \\ \hline
\end{tabular}

\caption{Simulated Arrays.  For each array, each constituent telescope's SEFD is listed. Beside each array name, its index is listed, which we reference in simulated experiments. \label{table:arrays}}
\end{center}
\end{table*}

The Current 230 GHz array consists of only the James Clark Maxwell Telescope (JCMT) in Hawaii, the Submillimeter Telescope Observatory (SMTO) in Arizona, and 2 dishes of the Combined Array for Research in Millimeter-wave Astronomy (CARMA) in California.  This array, which has been used in EHT campaigns for the last three years, has relatively poor sensitivity and little coverage in the North-South direction.  The next array, titled Precision, preserves the same baselines while increasing sensitivity.  The Submillimeter Array (SMA) replaces the JCMT, both of which are located in Hawaii, and the number of CARMA dishes is increased to 8.  Additionally, the bandwidth on all telescopes is increased from 1024 MHz to 4096 MHz, directly improving sensitivity by a factor of two.  In the Coverage array, the Large Millimeter Telescope (LMT) located in Mexico and the Atacama Pathfinder EXperiment (APEX) are added to provide better (u,v) coverage, especially in the North-South direction. APEX is currently being used in EHT campaigns, while efforts are underway to use the LMT at 230 GHz. Finally, the Complete array includes 50 dishes of the Atacama Large Millimeter Array (ALMA) in Chile, the Plateau de Bure Interferometer (PdBI) in France, the 30-meter telescope located atop the Pico Veleta (PV) in Spain, and the South Pole Telescope (SPT).  Instrumentation is also upgraded at the LMT, greatly increasing its sensitivity.  This Phase should be reached in the near future, once the ALMA phasing project \citep{fis13} is complete. 

So far, no data have been taken at 345 GHz.  Our Basic 345 GHz array consists of SMTO, SMA, APEX and the Atacama Submillimeter Telescope Experiment (ASTE).  The Coverage array then includes both 10 ALMA dishes, which provides much better sensitivity but no additional baselines, and PdBI, which provides an additional baseline at the edge of the resolution-limit due to turbulence.  Finally, the Future array includes 8 CARMA dishes, providing previously-lacking coverage on small baselines.  Such a short baseline as provided by CARMA-SMTO is essential to fitting models to 345 GHz data: before its introduction, there are no short baselines to determine the large-scale structure of the target.  The LMT can be added alternatively/in addition for a similar effect.  This is discussed further in section \ref{sec:constraints}.

\begin{figure*}
\begin{center}$
\begin{array}{cc}
\includegraphics[width=0.45\textwidth]{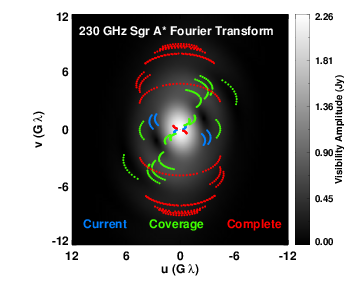} & 
\includegraphics[width=0.45\textwidth]{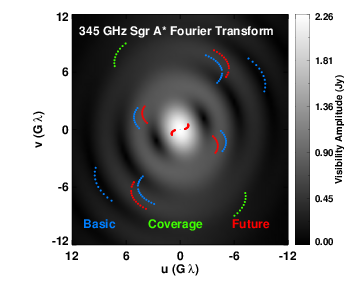} \\
\includegraphics[width=0.45\textwidth]{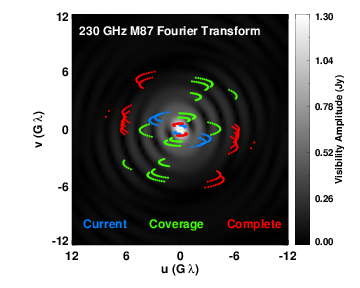} &
\includegraphics[width=0.45\textwidth]{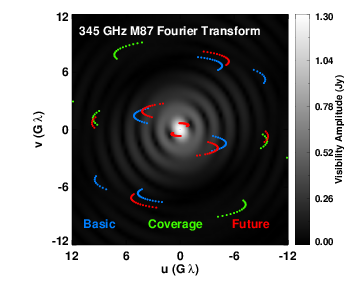} \\
\end{array} 
\newline
\begin{array}{ccc}
\includegraphics[width=0.33\textwidth]{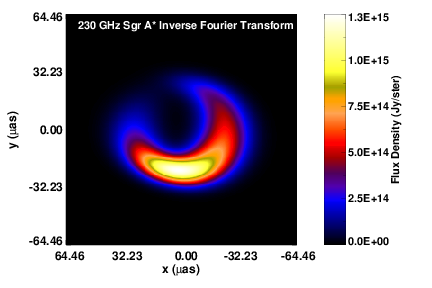} &
\includegraphics[width=0.33\textwidth]{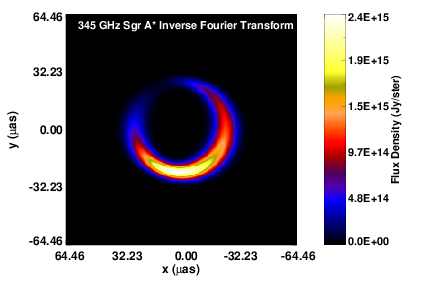} &
\includegraphics[width=0.33\textwidth]{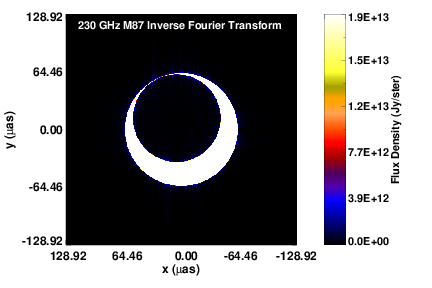}\\
\end{array}$
\caption{Crescent models and their Fourier Transforms.  The parameters used for these images are the best fit parameters to available 230 GHz data for Sgr A* and M87, listed in table \ref{table:best_fit_parameters}. {\it Above:}  The tracks traced out by different baselines have been over-plotted on our best fit ringless crescent models and color-coded according to the stages at which they enter our simulated arrays.  {\it Below:}  Inverse Fourier Transform images of the best fit crescent + ring model of Sgr A* and crescent model to M87 are shown.  A Hann window was applied in taking the Inverse Fourier Transforms to minimize edge effects.  The ring component is a small perturbation to the Sgr A* images, hidden underneath the much brighter crescent component.  Blurring is noticeably more severe at 230 GHz compared to 345 GHz.  Note that the best fit M87 crescent is considerably larger than that of Sgr A*.  Our best-fit images appear rotated with respect to figure \ref{fig:sim_images} due to the $180^\circ$ degeneracy present without phase information.  \label{fig:crescent_model}}
\end{center}
\end{figure*}

Figure \ref{fig:crescent_model} illustrates the (u,v) coverage of the EHT at both targets and frequencies, as well as the image of the underlying best-fit crescent model of each target at 230 GHz.  Observing Sgr A* at 230 GHz, very long baselines drawn from the south pole to the Northern Hemisphere are readily apparent, but the scales they probe are largely blurred away by electron scattering.  Moving to 345 GHz, one can observe the effects of diminished scattering, as well as the widening of all of the baselines at higher frequency.  Until CARMA is added to the 345 GHz array, there are no short baselines to gain information about the large-scale structure of either target.  The EHT has balanced (u,v) coverage in line-of-sight to M87, and there is no evidence for strong interstellar scattering along this line of sight.

\begin{figure*}
\begin{center}$
\begin{array}{cc}
\includegraphics[width=0.5\textwidth]{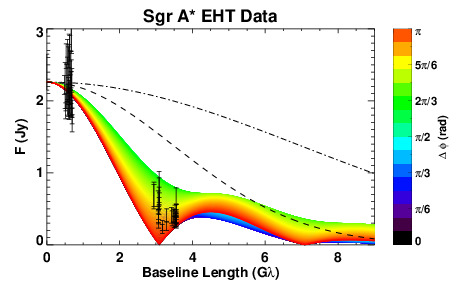} & 
\includegraphics[width=0.5\textwidth]{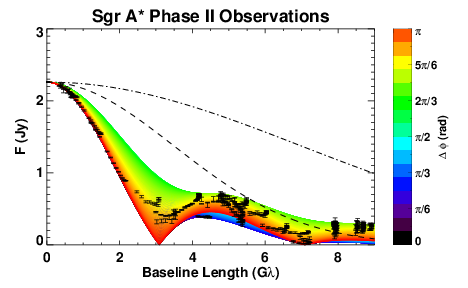} \\
\end{array}$
\caption{Precision and coverage of the current data and future data from the Complete 230 GHz array observing Sgr A*.   Real (left) and simulated (right) data points are overplotted upon the best-fit crescent + ring model to Sgr A*.  The colors from purple to red represent $\theta \in [\phi + \pi/2, \phi + 3\pi/2)$.  The dashed and dotted-dashed lines represent the scattering Gaussian along its minor and major axes respectively.  Although interstellar scattering suppresses flux at long baselines, the precision of the Complete EHT ensures that they can nevertheless make meaningful measurements. \label{fig:todayandComplete}}
\end{center}
\end{figure*}

Figure \ref{fig:todayandComplete} provides a visualization of the increased coverage and precision of the completed EHT compared to our current dataset.  Here, real or simulated data points are overplotted upon our blurred best-fit crescent + ring model to Sgr A*.  The colors from purple to red represent $\theta \in [\phi + \pi/2, \phi + 3\pi/2)$ rather than $\theta \in [\phi + \pi/2, \phi+\pi)$ as in figure \ref{fig:model_amplitudes}, since the asymmetry from the scattering Gaussian becomes noticeable at longer baselines.  For clarity, only errors bars are shown.  The dashed and dotted-dashed lines represent the scattering Gaussian along its minor and major axes respectively.  From this figure, it is clear that the Complete array will be capable of making precise detections at even its longest baselines, in spite of the presence of interstellar scattering.  These long baselines will therefore be useful in probing small-scale structure in our targets, which may arise e.g. from turbulence driven by accretion disk instabilities.

\section{Simulated Experiments}
\label{sec:experiments}

We perform a series of simulated EHT observations using models from \S\ref{sec:modeling} and mm-VLBI arrays from \S\ref{sec:arrays}, in order to i) assess its potential for detecting signatures of strong gravity and test black hole accretion theory; and to ii) determine which technical advances are most important for achieving these goals.

\subsection{Detecting the Photon Ring \label{sec:photon_ring}}

We do this by simulating observations of a model including a photon ring component, and asking when fitting models without this component is no longer statistically acceptable, e.g., when the fit would improve significantly from the inclusion of a photon ring.  We first generate 7 crescent models with photon ring flux contributions that range from 1\% to 40\% of the total flux, sampled evenly in logarithmic space. At 230 GHz, the total flux and geometric parameters are fixed at the best fit \emph{ringless} crescent parameters to current data of the target source, either Sgr A* or M87.  At 345 GHz, where there have been no observations, these parameters are fixed instead to the best fit ringless crescent parameters from simulation images, listed in table \ref{table:best_fit_parameters_sims}.  For Sgr A*, we use simulation MBQ, while for M87 we use model DJ1, both of which resemble crescents by eye and were simulated for each of the respective targets (Table \ref{table:sim_images} and Figure \ref{fig:sim_images}).

\begin{table*}
\begin{center}
\resizebox{\textwidth}{!}{
\begin{tabular}{ |c|c||cccccc| }
\hline
Object
   & Model
   & $R_{out}$ ($\mu$as)
   & $R_{in}/R_{out}$
   & $\phi$ ($^{\circ}$)
   & $F_C$ (Jy)
   & $\xi$
   & $F_R$ (Jy) \\
\hline
MBQ
   & Gaussian
   & $25.3^{+0.4}_{-0.4}$
   & $0.720^{+0.012}_{-0.014}$
   & $-71.2^{+1.9}_{-2.1}$
   & $1.705^{+0.003}_{-0.003}$
   & $-$
   & $-$ \\
MBQ
   & Crescent
   & $32.1^{+0.04}_{-0.18}$
   & $0.7026^{+0.0023}_{-0.0014}$
   & $-120.4^{+0.6}_{-0.7}$
   & $1.662^{+0.004}_{-0.016}$
   & $0.99997^{+0.00002}_{-0.00114}$
   & $-$ \\
DJ1
   & Gaussian
   & $26.7^{+0.5}_{-0.4}$
   & $0.576^{+0.011}_{-0.013}$
   & $-61.3^{+0.6}_{-0.6}$
   & $1.867^{+0.003}_{-0.003}$
   & $-$
   & $-$ \\
DJ1
   & Crescent
   & $27.51^{+0.05}_{-0.05}$
   & $0.6258^{+0.0014}_{-0.0104}$
   & $-90.7^{+0.6}_{-0.6}$
   & $1.828^{+0.004}_{-0.004}$
   & $0.99997^{+0.00002}_{-0.00027}$
   & $-$ \\
\hline
\end{tabular}
}
\end{center}
\caption{Best fit parameters to 345 GHz simulations of MBQ and DJ1, used as proxies for Sgr A* and M87 respectively.  These fits were performed by sampling the images with the Future 345 GHz array.  \label{table:best_fit_parameters_sims}}
\end{table*}

For each fraction of ring flux, each array, each frequency, and each target we find the best fit ringless crescent using the MCMC methods described in \S\ref{sec:modeling} and record the minimum $\chi^2$ value.  To minimize computational time, we use only 200 trials, which we found to reliably converge to the minimum $\chi^2$ value. To represent the ability to run multiple observing campaigns, we simulate observations and model fitting three times and find the \emph{maximum} value of $\chi^2$ from each of the three runs.  The maximum $\chi^2$ represents the case in which a ringless crescent fits the visibilities of a ringed crescent most poorly.  These $\chi^2$ values are used in order to determine the probability of a false negative:  the probability of concluding that a ringless crescent is consistent with the ringed crescent data. We do not account for possible false positives, which could be caused by degeneracies between different models.

In the case of the Gaussian distributed amplitude measurements, we can calculate the false negative probability using the CDF of the $\chi^2$ distribution:

\begin{center}
\begin{equation}
P_{FN} = 1 - \frac{\gamma(k/2,\chi^2/2)}{\Gamma(k/2)}  \label{eqn:P_FN},
\end{equation}
\end{center}

\noindent where $k = n-5$ and $n$ is the number of observed data points.

Closure phases are not Gaussian distributed, and so the $\chi^2$ values derived from the usual formula do not follow the chi-squared distribution.  In fact, the reduced $\chi^2$ value for the best fit model when measuring amplitudes and closure phases is not always 1, sometimes varying from $\sim0.5-1.5$.  This is most evident in the case of few closure triangles and low signal-to-noise.  Since the true PDF of $\chi^2$ values can be characteristically wider or thinner than its standard deviation would imply, this is not surprising.  Since our analysis hinges on correctly representing the tail of the distribution of $\chi^2$, we make a separate estimate of the CDF of $\chi^2$ values.  We simply ``brute-force'' sample the distribution by repeatedly calculating $\chi^2$ values of models fit to themselves.  We generate a large $\chi^2$ sample by simulating observations of the ``true'' image and calculating $\chi^2$ when it is fit with its own model.  This is computationally intensive, since the distribution is a function of both the instantiation of the EHT array and the model image.  The array determines which closure phases are measured, while the model image affects the signal-to-noise ratio, an input to equation \ref{eqn:closure_phase_error}.  We calculate 20000 values of $\chi^2$ for each image and array for which a closure phase analysis is done, giving us the true distribution of $\chi^2$.  This number of samples allows us to determine up to $\sim4 \sigma$ with some accuracy.  Rejection probabilities for test models are then calculated by determining where its $\chi^2$ value ranks in this probability distribution of $\chi^2$; we compute how likely it would be for the test-model's value of $\chi^2$ to be drawn from the distribution of truth models.

\begin{figure*}
\begin{center}$
\begin{array}{ccc}
\includegraphics[width=0.3\textwidth]{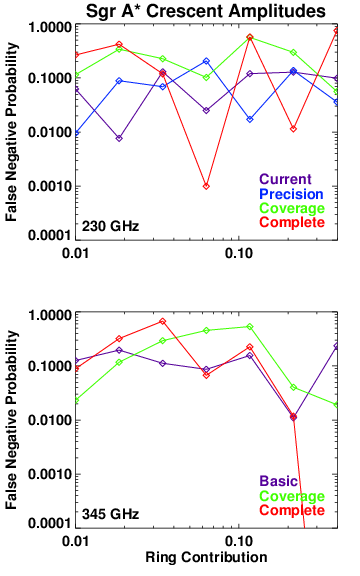} &
\includegraphics[width=0.3\textwidth]{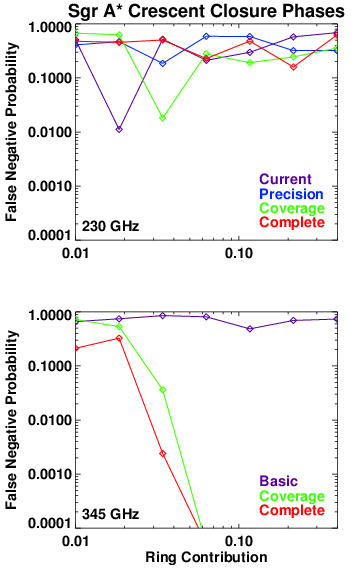} &
\includegraphics[width=0.3\textwidth]{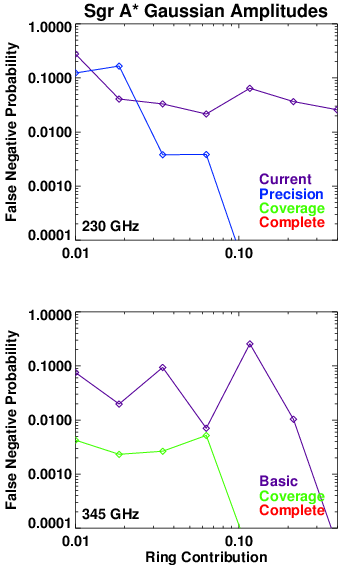}
\end{array} 
\newline
\begin{array}{ccc}
\includegraphics[width=0.3\textwidth]{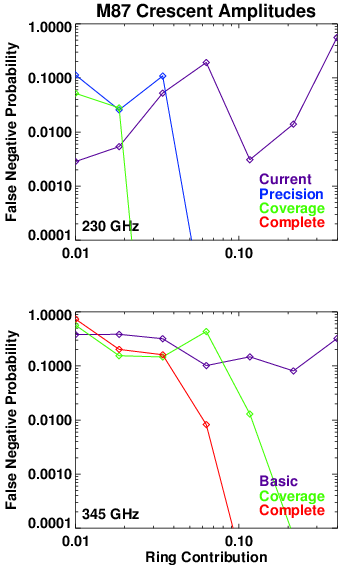} &
\includegraphics[width=0.3\textwidth]{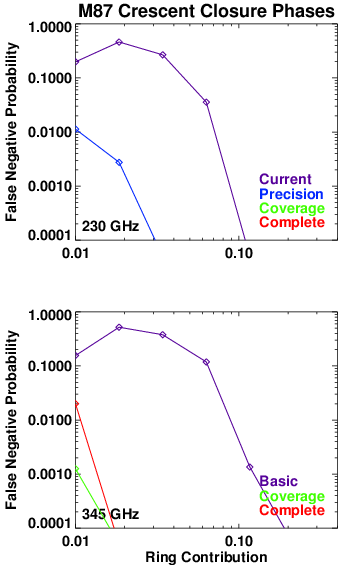} &
\includegraphics[width=0.3\textwidth]{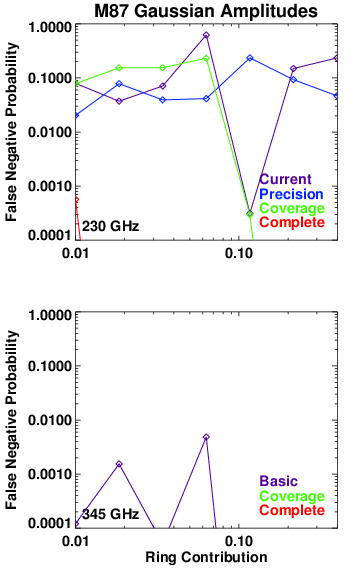}
\end{array}$
\caption{False negative probabilities as a function of ring contribution for the different scenarios.  Each track is color-coded based on the array used.  Local fluctuations can be attributed to differences in the random noise added to the visibilities in each run. Lines are not shown for cases where the false negative probability $p < 10^{-4}$ for the full range of photon ring flux.
\label{fig:ring_detection}}
\end{center}
\end{figure*}

Figure \ref{fig:ring_detection} displays the false negative probabilities as a function of array and ring contribution for a variety of experiments.  The most pessimistic results originate from Sgr A* at 230 GHz, where in all cases, the photon ring is undetectable even with the most advanced arrays and the greatest ring contributions.  However, this particular case suffers three problems:  (1) a crescent model can be degenerate with a ring model, since both have Bessel functions as their Fourier transforms; (2) the best fit ringless crescent to Sgr A* happens to subtend the same angular size as the photon ring ($\sim 27$ $\mu$as), placing them on top of one another; and (3) interstellar scattering suppresses flux on long baselines where the Bessel functions differ significantly. By repeating this experiment for Sgr A* without accounting for scattering, we determine that it is indeed the intrinsic geometry of the observed image that creates these pessimistic results (problems 1 and 2), rather than the presence of scattering (problem 3).  This is consistent with our result that moving to 345 GHz does little to rectify this problem, although the Future array manages to detect the photon ring when its flux is the brightest.  This geometry therefore represents the worst-case scenario. 

The most significant difference between a ring and a crescent in Fourier space is a phase asymmetry that corresponds to the offset of the crescent's hole from the image's center.  In addition, closure phases can be measured with much greater precision than amplitudes.  Hence, we see a marked improvement in our results by including closure phase information.  The most optimistic scenario for the Sgr A* crescents is one in which amplitudes and closure phases are measured at 345 GHz.  

These results are model dependent:  The similarity of a crescent to a ring makes ring detection particularly difficult.  We include for comparison results where the crescent in the observed image is substituted with the best fit Gaussians to Sgr A* and M87 (or their respective simulations at 345 GHz).  Here, even without phase information, all arrays aside from the current 230 GHz array are capable of detecting a ring with some amount of flux.  In fact many arrays fall off the plot, immediately distinguishing even a 1\% ring flux beyond 4$\sigma$ significance.  This is because Gaussian models provide little power on long baselines.  In this setup, all of the flux on those baselines must originate from the photon ring.

Results are also more optimistic if the source is changed to M87 (Figure \ref{fig:ring_detection}).  With a significantly larger best fit crescent size than photon ring size, model degeneracies are less severe for the M87 crescent.  This, combined with the absence of astrophysical noise, allows for ring detection at high significance even with amplitudes alone.  Advancing to closure phases, the arrays reach our $1/20000$ limit quite easily in almost all cases.  However, M87 Gaussian results happen to be more pessimistic than those for Sgr A*, due to the fact that the Gaussian's smaller intrinsic size does provide power at some of the longer baselines.

If the models used represent the true images of M87 and Sgr A*, a photon ring is more likely to be detected either by observing at 345 GHz or by observing M87. Yet in practice, both the model parameters and general shapes are not certain enough to warrant this conclusion. Using both sources and both observing frequencies provides the best opportunity for the EHT to detect a photon ring given possible degeneracies. In addition, a time-variability test, such as that developed in section \ref{sec:time_variability}, could be more effective in this regard, since the photon ring stays fixed in radius while the accretion flow can undergo significant structural variations.

\subsection{Differentiating Single Accretion Flow Images \label{sec:sim_fits}}

There are a variety of dynamical models that all provide good fits to existing mm-VLBI Sgr A* data at 230 GHz.  As the EHT develops, we will be able to distinguish these models. We illustrate this using some of the best fit simulation images from \S\ref{sec:modeling}.  We conduct an experiment to determine whether the EHT at various stages can distinguish between the best fit frames of time-dependent images calculated from different magnetohydrodynamical models of Sgr A* (Table \ref{table:sim_images}).

For each of these images, we calculate its Fourier Transform and interpolate onto the $(u,v)$ coordinates observed by a given array to obtain observed visibilities, with noise and errors applied as in \S\ref{sec:modeling}.  The fluxes of the M87 images are normalized to the values of our best-fit crescent model listed in table \ref{table:best_fit_parameters}, but we do not perform this step for Sgr A* best-fit images.  To obtain $\chi^2$, we then interpolate each model's Fourier Transform onto the model visibilities of the truth image.  Then, we calculate the number of $\sigma$ with which each observation can rule out the fitted model.  For amplitude measurements, we compare to the $\chi^2$ distribution and the built-in IDL function \verb&gauss_cvf& to obtain the number of $\sigma$ with which we can determine that these models are dissimilar.  For amplitude + closure phase measurements, we compute a sample of 20000 $\chi^2$ measurements and compare to this distribution rather than the canonical $\chi^2$ distribution as in \S\ref{sec:modeling}.

Table \ref{table:Sim_Fits} displays the arrays which are capable of distinguishing two models by $4\sigma$ or greater given different modes of observation.  We find that all models can be distinguished from one another with 4$\sigma$ significance prior to the inclusion of ALMA into the EHT.  For Sgr A*, the extra sensitivity provided by the Precision array is sufficient to make two of the distinctions, while the most difficult distinction is accomplished with the additional telescopes of the Coverage array.  For M87, the disk and jet models (DJ1 and J2 respectively) are distinguishable even with the Current array, implying that the current data may already be capable of constraining properties of its accretion flow.  Finally, the inclusion of closure phases allows even the Current array to make any distinctions.

\begin{table*}
\begin{center}
\begin{tabular}{ |c|c|c|c|c|c|c| }
\cline{2-7}
\multicolumn{1}{r}{} & \multicolumn{5}{|c|}{Sgr A*} & \multicolumn{1}{|c|}{M87} \\
\hline
Measurements & 90h vs. MBD & MBD vs. 915h & MBD vs. MBQ & MBD vs. 915h (345 GHz) & MBD vs. MBQ (345 GHz) & DJ1 vs. J2 \\
\hline
Amplitudes & Coverage & Precision & Precision & Basic & Basic & Current \\
Closure Phases  & Current & Current & Current & Basic & Basic & Current \\
\hline
\end{tabular}

\caption{Distinguishing three models for Sgr A* at 230 GHz.  The least advanced arrays capable of distinguishing the models to 4$\sigma$ significance are listed.}
\label{table:Sim_Fits}
\end{center}
\end{table*}

These results imply that additional data with the current array may already be able to provide new constraints on the accretion flows of Sgr A* and M87.  In practice, however, the inherent time variability of these models combined with parameter degeneracies make these comparisons less straightforward.

\subsection{Constraints on Crescent Parameters} \label{sec:constraints}

The previous subsection showed that the EHT can already begin to distinguish between viable theoretical models. In order to quantitatively measure the ability of the different instantiations of the EHT to constrain accretion models, we perform an experiment to determine how well each of the arrays could constrain crescent model parameters. We select the crescent model with the best fit parameters to Sgr A* as the observed image.  To isolate array performance, we use these same parameters at 345 GHz as well as 230 GHz.  As in \S\ref{sec:fitting}, we fit crescent models using MCMC, where the number of walkers increased to 300 in order to ensure that the posterior distribution was well-sampled.  We then calculate the uncertainties on these parameters by enclosing the 99.7\% of the models ($3 \sigma$) on either side of the mode of the posterior distribution, and averaging the uncertainty of either side.  

\begin{figure*}
\begin{center}$
\begin{array}{cc}
\includegraphics[width=0.45\textwidth]{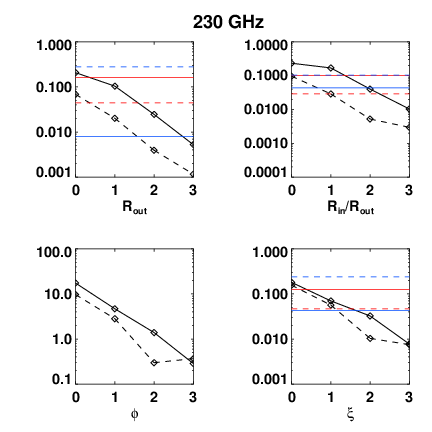} &
\includegraphics[width=0.45\textwidth]{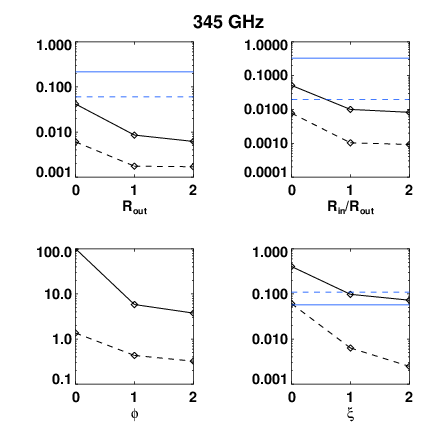}
\end{array}$
\caption{Fractional 3$\sigma$ uncertainties as a function of array for Sgr A*.  The black solid line indicates that measurements were made of amplitudes only, while the black dashed line indicates that closure phases were also measured.  The left two columns apply to 230 GHz arrays, while the right two columns apply to 345 GHz arrays.  Differences between equivalent crescent parameters for MBD vs. MBQ and MBD vs. 915h are represented by blue solid and dashed lines.  In the 230 GHz results, we overplot the RMS fluctuations in equivalent crescents to simulations 515h (red solid) and 90h (red dashed).  The x-axis displays the index of the array in question, instead of its name.
\label{fig:constraints}}
\end{center}
\end{figure*}

Figure \ref{fig:constraints} displays the results of this test.  For the parameter $R_{out}$, uncertainties were normalized by dividing by the mean value of each of those parameters.  Uncertainties in $\phi$ are in units of degrees.  Solid black lines represent runs where measurements are made using amplitudes only, while dashed blacklines represent runs involving both amplitudes and closure phases.  For the 230 GHz plots, the red solid and dashed lines correspond to the RMS variations of the best fit parameters to simulations 515h and 90h, one of the most and least varying simulations respectively.  Fluctuations between different frames of 515h are just at the threshold necessary to detect time-variability with the Current array, and this is later developed in section \ref{sec:time_variability}.  The blue solid and dashed lines represent the differences in best-fit crescent model parameters for MBD vs. MBQ and MBD vs. 915h, two of the most similar and least similar models respectively.  Interestingly, while the differences between MBD and MBQ are difficult to determine at 230 GHz, they become less similar at 345 GHz. 

The progression from amplitude to closure phase measurements is more pronounced at 345 GHz than at 230 GHz, due to increased signal-to-noise on long baselines in the absence of electron scattering.  The plateau in closure phase constraints in the last two 230 GHz arrays, most evident in $\phi$ and $\xi$, suggest that a limit is reached due to electron scattering.

\subsubsection{Possible Large Scale Problems at 345 GHz \label{sec:345ghz_large}}

\begin{figure*}
\begin{center}
\includegraphics[width=0.9\textwidth]{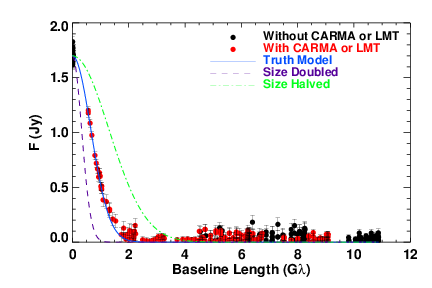}
\caption{The necessity of short baselines at 345 GHz.  The Coverage 345 GHz array fails to constrain the size of a symmetric Gaussian.  The truth image has a radius of 50 $\mu$as, but without the inclusion of the red data points provided by CARMA or LMT on smaller scales, models with both twice and half the true size can reproduce the data.
\label{fig:large_scale_problems}}
\end{center}
\end{figure*}

A subtle issue that we encounter is that the 345 GHz arrays poorly constrain large-scale ($\gtrsim 30 \, \mu$as) structure without the inclusion of short baselines.  This is not evident in the previous test (and those to come), since crescent and ring models both have power at high spatial frequencies, and can therefore be constrained using only large baselines.  Figure \ref{fig:large_scale_problems} demonstrates this shortcoming.  Here, we simulate observations of a symmetric Gaussian 50 $\mu$as in radius, with a flux of 1.7 Jy, a reasonable flux taken from simulation MBQ.  We compare data points obtained using the Coverage array with data points obtained using the Future array, with the addition of the LMT, another telescope capable of providing shorter baselines.  Data points from baselines which include CARMA or the LMT are plotted in red, while those from baselines without either are plotted in black.  To calculate error bars, the SEFD of the LMT is taken to be 13700, as in \citet{lu14}.  Model visibilities of Gaussians with the true radius, half the radius, and twice the radius are overplotted.  It is only with the inclusion of CARMA that the true size can be constrained--all other baselines either measure the total flux, or resolve out the source.

This problem should be exacerbated on even larger scales, and it is not uncommon for simulated accretion flows to subtend areas ~2-3 times larger.  Hence, caution should be used interpreting future data obtained with the EHT's 345 GHz arrays if shorter baselines are not acquired.  It should be noted that even our Future array does not constrain large-scale structure in the North-South direction.  Obtaining 345 GHz capabilities at additional telescopes such as the LMT help could rectify this problem.  However, even without short baselines, 345 GHz data may still be quite valuable for testing models in conjunction with 230 GHz observations. For example, current viable models make significantly different predictions for even the total flux at 345 GHz \citep[e.g.,][]{dex10}.

\subsection{Structural Variability On Event Horizon Scales\label{sec:time_variability}}

Both Sgr A* and M87 exhibit significant ($\simeq 50\%$) variability on relevant timescales for EHT observations: $\sim$ hr for Sgr A*, $\sim$ yr for M87, and the simulated images we use are also time-variable at consistent amplitudes. \citet{fis11} found no significant evidence of structural variability in Sgr A* data even as the total flux changed by $\simeq 30\%$ between observations, but it should be readily detected with future data. Here we assess the sensitivity of the EHT to structural variability as predicted by the models. We answer this question using both time-dependent accretion flow and jet images calculated from MHD simulations, and geometric crescents with rings, which make fewer assumptions about the physical environment around the black hole.  For each test, we assume three different days of observation.  The model crescent or simulation image is static during each individual day, but no two days' images are alike. This assumption is justified for M87, where the light-crossing time of the event horizon is $\gtrsim 1$ day. Sgr A* varies on hour timescales \citep[e.g.,][]{dex14}. Our simulated days of observation could also represent separate segments within a single day for the case of Sgr A*.

\subsubsection{Crescent Models}
Geometric crescent models are computationally cheap to generate, and their variability is easier to quantify.  We consider variability between days in each model parameter except total flux of 1\%, 2\%, and 5\% through 25\% with 5\% steps, levels consistent with predictions of MHD simulations (see Table \ref{table:real_fluctuations}).  We begin with the best fit crescent parameters to Sgr A* or M87 given by table \ref{table:best_fit_parameters} at 230 GHz, or the best fit crescents to simulations MBQ and DJ1 given by table \ref{table:best_fit_parameters_sims}.  Then, we multiply each geometric parameter by $1 + x$ for each day, where $x$ is a random variable drawn from a Gaussian distribution with a mean of 0 and a standard deviation equal to the value of the variability in question.  Since variations in the total flux of the image are trivial to deduce from data, we keep the fluxes of the crescent fixed, so that we only explore variations in the spatial parameters of the crescent model.

We then use MCMC to fit to a combined dataset of three days of observation in two different ways:  First, a static fit in which on model must fit all three days of observation, and secondly a time-varying fit in which parameters are allowed to vary between different days.  In order to determine whether a static or time-varying fit is statistically favored, we record the best value of $\chi^2$ in each case and determine whether it is consistent with the expected distribution of $\chi^2$ values, given the number of data points and free parameters.  When we include closure phase measurements, non-Gaussian statistics impel us to ``brute-force'' sample the distribution directly.  As in  \S\ref{sec:photon_ring}, we calculate 20000 samples of the distribution of $\chi^2$ values with which to compare our data. 

\begin{table*}
\begin{center}
\resizebox{\textwidth}{!}{
\begin{tabular}{ |c|c|c||ccccccc| }
\hline
Source & Frequency & Measurements & 1\% & 2\% & 5\% & 10\% & 15\% & 20\% & 25\% \\ \hline
Sgr A* & 230 GHz  & Amplitudes & Complete & Coverage & Precision & Current & Precision & Precision & Current \\
Sgr A* & 230 GHz  & Closure Phases & Precision & Current & Current & Current & Current & Current & Current \\
Sgr A* & 345 GHz  & Amplitudes & Never & Coverage & Coverage & Basic & Basic & Basic & Basic \\
Sgr A* & 345 GHz  & Closure Phases & Basic & Basic & Basic & Basic & Basic & Basic & Basic \\
M87 & 230 GHz  & Amplitudes & Complete & Precision & Coverage & Precision & Current & Current & Current \\
M87 & 230 GHz  & Closure Phases & Current & Current & Current & Current & Current & Current & Current \\
M87 & 345 GHz  & Amplitudes & Future & Basic & Coverage & Basic & Basic & Basic & Basic \\
M87 & 345 GHz  & Closure Phases & Basic & Basic & Basic & Basic & Basic & Basic & Basic \\
\hline
\end{tabular}
}
\caption{Detecting time variability with crescent models.  The array which successfully distinguishes time-variability to 4$\sigma$ is listed.
\label{table:time_synthetic}}
\end{center}
\end{table*}

Table \ref{table:time_synthetic} shows the results of this experiment, displaying the first array which detects time variability at the $4 \sigma$ level for different levels of fluctuation.  We notice no preferred source for this case.    By performing a test in which we repeat this experiment for M87 with its predicted flux doubled (to a 230 GHz flux more similar to Sgr A*), we determine that these results are not limited by the signal-to-noise ratio.  After all, we expect these large-scale fluctuations to be detectable by the short baselines with the highest signal-to-noise.  Neither do we see greater success rates at 345 GHz versus 230 GHz.  One would hope to perform this test at both frequencies, however.  First, the sources may have different fluxes at each frequency; neither source has been resolved at 345 GHz, so this remains an open question.  Second, observations at 345 GHz plumb a deeper optical, and therefore physical, depth and may therefore undergo larger amplitude structural variations. The factor which has the greatest effect on detecting time-variability is the inclusion of closure phase information.  As discussed in \S\ref{sec:closure_phases}, longer integration times will allow more precise measurements of closure phases than visibility amplitudes.

For comparison with physical models, we fit crescent models to each frame of various simulation images, and record the best fit parameters.  A mesh in (u,v) space is generated, linearly sampling with 30 steps in each parameter $\sqrt{u^2+v^2}$ from 0 to 5 G$\lambda$, and $\phi$ from 0 to $\pi$.  Physical models and crescent models are then fit using MCMC with visibility amplitudes obtained at these data points.  Data points are obtained by this method rather than by simulating observations with a real array because (i) this ensures an even sampling of (u,v) space, with equally-weighted points and (ii) crescent models are not expected to fit data points where $\sqrt{u^2+v^2} \gtrsim 5 \, \mathrm{ G}\lambda$, where power is dominated by small scale structure.  Given these fits, we compute the standard deviation in each parameter and divide by the average to obtain fractional time variability for each model.  These are listed in table \ref{table:real_fluctuations}.  Since most of the model's crescents vary much more than the threshold needed to detect time variability for the Current array, we predict that geometric time variability should be detectable in EHT data once coherently averaged closure phases become available.  In the following section, we perform a more rigorous experiment to lead to this conclusion.

\begin{table}
\begin{center}
\begin{tabular}{|c|cccc|}
\hline
Model & $R_{out}$ ($\mu$as) & $R_{in}/R_{out}$ & $\phi \, (^\circ)$ & $\xi$ \\ \hline
515h & 16\% & 27\% & 58\% & 14\% \\ 
90h & 4\% & 6\% & 10\% & 6\% \\ 
MBQ & 8\% & 15\% & 19\% & 10\% \\ 
DJ1 & 16\% & 19\% & 22\% & 27\% \\ 
J2 & 17\% & 15\% & 19\% & 10\% \\ 
\hline
\end{tabular}
 \caption{Best fit crescent variability in MHD accretion flow models of Sgr A* and M87.  After fitting crescents to each frame of the simulation, the standard deviation of each parameter was measured and divided by its average.  Fluctuations of this magnitude should be detectable by EHT data including closure phases.  DJ1 appears deceptively variable, but this is simply due to the fact that it is more symmetric than the other simulations, causing its crescent parameters (esp. $\phi$) to be more poorly constrained.  \label{table:real_fluctuations}}
 \end{center}
 \end{table}

\subsubsection{Simulated Models}

We perform a similar test to compare our results with time-varying crescents to physical simulated images. For a given simulation, we first calculate the $\chi^2$ obtained by fitting each of its images to each of its other images when observed with each of the different arrays.  Variations in flux alone from frame to frame are enough to make this experiment trivial for even the most basic arrays.  Hence, to isolate geometric fluctuations, we normalize the flux in each frame to our current best-fit fluxes of each target.  

We determine the fraction of the time that each of the arrays can detect time-variability by simulating 20000 observing campaigns, each of which contains observations for three different epochs.  For each of these epochs, we assume the source remains fixed at a random frame from its simulation.  The best-fit time-varying model is simply given the three different frames themselves, while the best-fit static model is the single frame which minimizes the total $\chi^2$ when fit to all three epochs simultaneously.  Saving each of these $\chi^2$ values for each campaign, we obtain the distribution of $\chi^2$ that we expect for both time-varying and static fits.

Given these distributions, we then calculate the percentage of the time that the static fits are more than $4\sigma$ inconsistent with the time-varying fits.  In table \ref{table:sim_time_detection}, we display this fraction for each image, each array, and each mode of observation.  We find that it is only the current 230 GHz array that has an appreciable difficulty in detecting time-variability.  The Precision array, which has the same baselines as the Current array but a more sensitive Hawaii telescope, additional CARMA dishes, and higher bandwidth, performs dramatically better than the Current array.  This suggests that precision, rather than (u,v) coverage, is the limiting factor in determining time-variability, as it occurs at all spatial scales plumbed by the EHT.  Once again, the inclusion of closure phases also boosts the chances of distinguishing models significantly.  

\begin{table*} 
\begin{center}
\begin{tabular}{ |c|c|c|c|c|c|c|c|c| }
\cline{3-9}
\multicolumn{2}{r}{}  & \multicolumn{4}{ |c| }{230 GHz} & \multicolumn{3}{ |c| }{345 GHz} \\ 
\hline 
Measurements & Simulation & Current & Precision & Coverage & Complete & Basic & Coverage & Future \\ \hline

\multirow{5}{*}{Amplitudes} 
& 515h & 72\% & 100\% & 100\% & 100\% & & & \\ 
& 90h & 6\% & 94\% & 100\% & 100\% & & &  \\ 
& MBQ & 49\% & 97\% & 100\% & 100\% & & & \\ 
& DJ1 & 3\% & 83\% & 93\% & 100\% & 84\% & 100\% & 100\% \\ 
& J2 & 44\% & 99\% & 100\% & 100\% & 42\% & 100\% & 100\% \\ \hline

\multirow{5}{*}{Closure Phases} 
& 515h & 100\% & 100\% & 100\% & 100\% & & & \\ 
& 90h & 98\% & 100\% & 100\% & 100\% & & &  \\ 
& MBQ & 99\% & 100\% & 100\% & 100\% & & & \\ 
& DJ1 & 90\% & 100\% & 100\% & 100\% & 100\% & 100\% & 100\% \\ 
& J2 & 100\% & 100\% & 100\% & 100\% & 100\% & 100\% & 100\% \\ \hline

\end{tabular}
\caption{Percentage of time-variability detection at 4$\sigma$ significance or greater.  For the DJ1 and J2 models, images have been calculated at the two different frequencies.}
\label{table:sim_time_detection}
\end{center}
\end{table*}

\section{Discussion}
\label{sec:discussion}

This work presents a concrete demonstration of the potential of the EHT to do transformative science within the framework of plausible observing parameters and models for the source structure in Sgr A* and M87. However, there are a number of limitations to the methods used here. The arrays we consider (Table \ref{table:arrays}) correspond roughly to the technical roadmap of the EHT experiment, but are necessarily approximations of the true evolution of its technological capabilities in time.  Future array developments which we have not considered in our analysis include the addition of the 12m Greenland Telescope \citep[GLT,][]{ino14}, which will provide long baselines for M87 observations, and the addition of LMT functionality at 345 GHz, which will help constrain large-scale structure at this frequency.

Simulating future observations requires assuming models for each source. We have made plausible choices based on best fit images from state of the art simulations, and a model-independent geometric crescent approximation aiming to capture the important relativistic effects. Both choices are first steps. The former models are subject to systematic uncertainties in the microphysics and initial conditions of the simulations, which should be explored in the future. The geometric crescent model is physically motivated and has more parameters (5) than asymmetric Gaussian (4) or ring (3) shapes, but is still too simplistic to reproduce the theoretical images in any detail. In addition, while we consider an azimuthally symmetric photon ring contribution for simplicity this feature is also asymmetric in theoretical models due to Doppler beaming. Analytic extensions including additional parameters \citep{ben13} are a promising means to extend these models to quantitatively fit future EHT data sets.  We have also separately carried out static and time-variable tests. The most effective method for detecting a photon ring may depend on its stationarity relative to the variable emitting region.

While beyond the scope of this paper, the EHT array will also be able to measure spatially resolved dual polarizations, which will provide independent constraints on the structure of magnetic fields and geometry of the emitting region.  Advantages of polarimetric interferometry include the immunity of fractional polarization to scatter-broadening, thereby circumventing electron turbulence in line of sight to Sgr A*, and the ability to reach greater astrometric precision than the size scales probed by an array's baselines.  Recent work by \citet{joh14} demonstrates the promise of polarimetric interferometry for achieving precise relative astrometry of a variable polarized image compact.

Similar to previous work \citep{bro11a}, we find that coherently averaged closure phases are a powerful observable for EHT science. Closure phase can now be measured, and increasing their sensitivity is an important future goal for the observations. Even with current capabilities, we predict that structural variability should be detectable with additional data. The amplitude and type of variability observed will provide a powerful constraint on accretion models, possibly including the direct detection of magnetic turbulence in a black hole accretion flow.

Here we only consider the detectability of a Schwarzschild photon ring. In principle, the characterization of the photon ring, particularly its shape and phase offset from the rest of the image, could allow a measurement of black hole spin \citep{tak04} and a test of the no-hair theorem \citep{joh10}. This can be explored with simulated EHT data in future work, either using static models as we have done here or by match filtering the expected feature in time-dependent simulated data.

The models we consider assume general relativity.  If the Kerr geometry is modified, it becomes difficult to disentangle signatures of modified gravity from different accretion flow geometries.  A computationally-expensive, full parameter space search is required, as recently explored by \citet{bro14}.

Despite these caveats, the techniques that we develop in this paper do not depend on the underlying models.  As new simulation models or analytic parameterizations are introduced, they can be used while applying the same methodology.

\section{Conclusion}
\label{sec:conclusion}

With both geometric crescents and hydrodynamical accretion models, we have devised and simulated a variety of tests that can be used on current and future EHT data and assessed the array's capabilities during its stages of development.  The main results of this paper are summarized as follows.

\begin{itemize}
\item The photon ring, a signature of Kerr strong gravity, may be detected by the Event Horizon Telescope once ALMA and LMT have joined the array (Complete) for its predicted brightness of $\simeq 1-10\%$ of the total flux (Section \ref{sec:photon_ring} and Figure \ref{fig:ring_detection}). 
\item The use of closure phase information greatly increases the constraining power of the EHT.  Due to the ability to coherently average closure phase information over time periods much longer than atmospheric coherence time, the increase in precision gained through measuring closure phases more than makes up for the lack of full phase information.
\item The EHT should be able to discriminate between many best-fit theoretical models from numerical simulations even with additional data from the current array, provided that closure phases can be robustly measured with high precision (Section \ref{sec:sim_fits}). Model constraints will become far stricter still in the next few years as bandwidth is increased and stations are added to the array (e.g., LMT and ALMA). EHT observations can therefore provide a stringent test of contemporary accretion theory and a precise measurement of black hole and accretion flow parameters.
\item EHT closure phases are also currently sensitive to structural variability at the level present in the models (Section \ref{sec:time_variability}). Current data may therefore allow the first detection of structural variability on event horizon scales around a black hole. Time-domain EHT measurements provide could allow the direct study of instabilities in accretion flows both on the scales of the entire image and its fine structure, e.g. as may result from MHD turbulence driven by the MRI (right panel of Figure \ref{fig:todayandComplete}).
\item The assumed underlying image model can significantly impact the results achievable by the EHT. Using the full complement of two sources (Sgr A* and M87) and frequencies (230 and 345 GHz) can mitigate possible degeneracies between the accretion flow or jet image and the photon ring (e.g., the difference between the results for 230 GHz M87 or 345 GHz Sgr A* and 230 GHz Sgr A* in Figure \ref{fig:ring_detection}). Using M87 and 345 GHz are particularly important to avoid the strong interstellar scattering towards Sgr A*, which can still limit the science capabilities of observations at 230 GHz.

\end{itemize}

\begin{table*}
\begin{center}
\resizebox{\textwidth}{!}{
\begin{tabular}{ccl}
\hline
Frequency & Array & Science Goals within Reach\\
\hline
230 GHz & Current & With amplitudes only, detect expected structural time variability across observation epochs for some physical   \\
& &  models and distinguish jet from disk models.  With closure phases, detect structural time variability for all physical    \\
& &  models, and break degeneracies between current best-fit models to Sgr A*.\\
 & Precision & With amplitudes only, detect structural time variability across observation epochs for all physical models,  \\
& & break degeneracies between most current best-fit models to Sgr A*, and distinguish M87 photon ring from crescent. \\
 & Coverage & With amplitudes only, break all degeneracies between current best-fit models to Sgr A* and probe structural \\
 & & time-variability at ~5\% levels. \\
 & Complete & Detect photon ring in Sgr A* and M87 for all levels of predicted flux for a wide range of models. \\
& & With amplitudes alone, probe structural time-variability at $\sim 1$\% levels. \\
 \hline
345 GHz & Basic & With amplitudes alone, place constraints on time-variability of physical models.  With closure phases, \\
& & detect this time-variability with certainty. \\
 & Coverage & With closure phases, distinguish Sgr A* photon ring from crescent. \\
 & Future & With amplitudes only, distinguish Sgr A* photon ring from crescent.  Constrain large scale accretion flow.\\
\hline
\end{tabular}
}
\end{center}
\caption{Major science goals attainable by each array.   \label{table:summary}}
\end{table*}

\section*{acknowledgements}
We thank S. Doeleman and E. Quataert for useful discussions. We are grateful to L. Benkevitch, S. Doeleman, V. Fish, and R. Lu for sharing the MAPS software and for assistance with its use.

\footnotesize{
\bibliography{ms}
}
\label{lastpage}

\end{document}